\documentclass{ifacconf}

\usepackage{amsmath,amssymb,amsfonts}
\usepackage{graphicx}      
\usepackage{natbib}        
\usepackage{booktabs}
\usepackage{makecell}
\usepackage{multirow}
\usepackage{url}
\usepackage[switch]{lineno}

\usepackage{soul}
\usepackage{color, xcolor}
\definecolor{lightblue}{RGB}{0, 191, 255}
\sethlcolor{lightblue}
\soulregister\cite7
\soulregister\eqref7
\soulregister\ref7

\begin{document}
	
\begin{frontmatter}
	
\vspace*{-1.2cm}
\begin{flushleft}
	\small Accepted by IFAC World Congress 2026
\end{flushleft}

\title{ASIND: Alternating Sparse Identification for Predicting Network Dynamics Without Knowledge} 

\thanks[footnoteinfo]{This work is supported by the National Natural Science Foundation of China (Grant No.~T2293771), the STI 2030-Major Projects (Grant No.~2022ZD0211400), the New Cornerstone Science Foundation through the XPLORER PRIZE, the USA National Science Foundation (Grant No.~2047488), the Rensselaer-IBM Future of Computing Research Collaboration, and the Jiangsu Provincial Scientific Research Center of Applied Mathematics (Grant No.~BK20233002). (Corresponding author: Linyuan L\"{u})}

\author[First]{Mingyu Kang} 
\author[Second]{Jianxi Gao} 
\author[Third]{Wenwu Yu}
\author[First]{Linyuan L\"{u}}

\address[First]{School of Cyber Science and Technology, University of Science and Technology of China, Hefei, Anhui 230026, China (e-mail: \{kangmingyu, linyuan.lv\}@ustc.edu.cn).}
\address[Second]{Department of Computer Science, Rensselaer Polytechnic Institute, Troy, New York 12180, USA (e-mail: gaoj8@rpi.edu)}
\address[Third]{School of Mathematics, Southeast University, Nanjing, Jiangsu 210096, China (wwyu@seu.edu.cn)}

\begin{abstract}                
Identifying network dynamics is a critical yet challenging task to to understand the mechanism of real-world social systems. There are two types of algorithms, and one requires the knowledge of self-dynamics function, interactive function, and interactive network to sparsely identify the network dynamics. Another one does not require any knowledge, but use simple functions to universally approximate complex functions. However, this type of algorithms lack interpretability, and the functional space is too extensive to search efficiently. Thus, to address this issue, this work proposes an Alternating Sparse Identification of Network Dynamics (ASIND) algorithm to sparsely identify the self-dynamics function, interactive function and interactive network alternatively. Extensive experiments are conducted to show the state-of-the-art identification and 100-steps prediction performance compared to the baseline. The experimental results also show the weak identifiability of interactive network, that means different networks can generate highly similar trajectories of network dynamics. \textbf{The code is available at \url{https://github.com/KMY-SEU/ASIND}}.
\end{abstract}


\end{frontmatter}

\section{Introduction}

Identifying network dynamics is a critical approach to understand the mechanism of real-world social systems. The network dynamics is defined as
\begin{equation}
	\dot{x_i} = F(x_i) + \sum_{j=1}^N A_{ij}G(x_i, x_j),
	\label{netdyns}
\end{equation}
where $x_i\in \mathbb{R}^d, i = 1, \dots, N$ are the $d$-dimensional activities of $N$ nodes. The function $F$ is the self-dynamics of node $i$, and the function $G$ is the interactive dynamics of node $i$ and node $j$. $A$ is the weighted adjacency matrix, and $A_{ij} \geq 0$. The interactive property is fundamental and widely existing in many natural physical systems~\citep{MM2016Gao, University2013NP, Kang2026, STBN2024, NHCE2024}. By understanding the underlying mechanism, one can further investigate the resilience and synchronization of social systems, predict long-term evaluation of social behaviors, and then guide social regulation~\citep{ERA2024, Yu2010}. 

To identify the network dynamics, \cite{TPI2022NCS} propose a two-phase inference algorithm. The algorithm assumes that the network $A$ is known, and then sparsely identifies the basis matrix of $F$ and $G$ that composed of two groups of basis functions. By contrast, \cite{TSP2022PNAS} propose a two-step prediction algorithm, that assumes $F$ and $G$ are known, but $A$ is unknown, which is exactly the opposite of the two-phase inference algorithm. Similarly, \cite{RulkovMap2023PRL} propose an algorithm to identify the network $A$ for dynamics of Rulkov maps. Thus, these algorithms must either know $F$ and $G$, or know $A$ in prior, but most natural dynamics do not come with such knowledge.

Then, another type of algorithm, e.g., SINDy~\citep{SINDy2016PNAS}, tries to sparsely identify the nonlinear dynamics without such knowledge. Then, \cite{PDE2017SA} expand that algorithm to the identification of partial differential equations. \cite{SBL2019NC} sparsely identify the nonlinear dynamics with Bayesian principle. Moreover, there are also some research works~\citep{NODE2018NIPS, cfc2022NMI, GNODE2023chaos} that uses artificial neural networks to universally approximate nonlinear dynamics, but the artificial neural networks lack interpretability for the system mechanism. And more importantly, this makes the functional space too extensive to search network dynamics efficiently.   

Thus, to identify the network dynamics without any knowledge, and then interpret the network system mechanism, Alternating Sparse Identification of Network Dynamics (ASIND) algorithm is proposed here. The proposed algorithm focuses on the sparse identification of network dynamics, and does not assume the known $F$, $G$ and $A$ in prior. It parameterizes $F$, $G$ and $A$ together, and identifies $A$ firstly, then identifies $F$, $G$ afterwards, and iterate this process until convergence.   

Thus, the main contributions of this work are as follows:
\begin{enumerate}
	\item[(i)] ASIND algorithm is proposed to identify network dynamics without the knowledge of $F$, $G$ and $A$.
	\item[(ii)] Experiments are conducted to validate the state-of-the-art performance on the network dynamics identification, compared to the popular SINDy algorithm~\citep{SINDy2016PNAS}. Moreover, the results also show that the network structure $A$ is weakly identifiable, that means different networks are likely to generate highly similar trajectories. 
\end{enumerate}

\section{Alternating Sparse Identification of Network Dynamics}

The sparse identification issue of network dynamics in Eq.~(\ref{netdyns}) can be defined as 
\begin{equation}
	\begin{aligned}
		&\min_{w, A} \sum_{i=1}^N \sum_{m=1}^{M_1 + M_2} \left\Vert w_{im}\right\Vert_1 + \sum_{i=1}^N \sum_{j=1}^N \left\Vert A_{ij}\right\Vert_1 \\
		\text{s.t.}~&\dot{x}_i = \sum_{m=1}^{M_1} w_{im} F_m(x_i) + \sum_{m=1}^{M_2} w_{im} \sum_{j=1}^N A_{ij}G_m(x_i, x_j), \\
		& A_{ij} \geq 0, i = 1, \dots, N, j = 1, \dots, N.
		\label{original}
	\end{aligned}
\end{equation}
Here, for each node $i$, $F_m, m = 1, \dots, M_1$ and $G_m, m = 1, \dots, M_2,$ are the basis function, and the $w_{im}, m = 1, \dots, M_1 + M_2$ are the corresponding coefficients. Moreover, $\left\Vert\cdot\right\Vert_1$ is the $l_1$-normalization operator that enables the $w_{im}$ and $A_{ij}$ to be sparse.

Then, Eq.~(\ref{original}) can be rewritten as 
\begin{equation}
	\begin{aligned}
		&\min_{w, A, \lambda} \mathcal{L}(w, A, \lambda) \\ 
		\text{s.t.}~&A_{ij} \geq 0, i = 1, \dots, N, j = 1, \dots, N.
		\label{new}
	\end{aligned}
\end{equation}
where, 
\begin{equation}
	\small
	\begin{aligned}
		&\mathcal{L}(w, A, \lambda) \\ 
		= &\sum_{i=1}^N\sum_{m=1}^{M_1 + M_2} ||w_{im}||_1 + \sum_{i=1}^N \sum_{j=1}^N ||A_{ij}||_1 \\ 
		& + \sum_{i=1}^N \lambda_i \left(\dot{x}_i - \sum_{m=1}^{M_1} w_{im} F_m(x_i) - \sum_{m=1}^{M_2} w_{im} \sum_{j=1}^N A_{ij}G_m(x_i, x_j)  \right) \\ 
		& + \frac{\rho}{2} \sum_{i=1}^N \left\Vert\dot{x}_i - \sum_{m=1}^{M_1} w_{im} F_m(x_i) - \sum_{m=1}^{M_2} w_{im} \sum_{j=1}^N A_{ij}G_m(x_i, x_j) \right\Vert_2^2.
	\end{aligned}
\end{equation}
Here, $\left\Vert\cdot\right\Vert_2$ is the $l_2$-normalization operator for augmented Lagrangian term. $\lambda_i$ is the Lagrange multiplier, and $\rho$ is the augmented Lagrangian penalty parameter. The Eq.~(\ref{original}) and Eq.~(\ref{new}) have the same optimal solution, and the Eq.~(\ref{new}) can be alternatively solved by the following steps:
\begin{equation}
	\begin{aligned}
		&A^{(k+1)} = \mathop{\arg\min}\limits_{A} \mathcal{L}\left(w^{(k)}, A, \lambda^{(k)}\right), \\
		\text{s.t.}~&A_{ij} \geq 0, i = 1, \dots, N, j = 1, \dots, N,
	\end{aligned}
	\label{Ak}
\end{equation}
and then,
\begin{equation}
	w^{(k+1)} = \mathop{\arg\min}\limits_{w} \mathcal{L}\left(w, A^{(k+1)}, \lambda^{(k)}\right),
	\label{wk}
\end{equation}
and then,
\begin{equation}
	\lambda^{(k+1)} = \lambda^{(k)} + \alpha \times \frac{\partial \mathcal{L}\left( w^{(k+1)}, A^{(k+1)}, \lambda \right)}{\partial \lambda},
	\label{lk}
\end{equation}
where $\alpha$ is the step size to control the gradient descent of $\lambda$.

Then, to update $A^{(k+1)}$ in Eq.~(\ref{Ak}), independently for $A_{i}^{(k+1)} = \left[ A_{i1}^{(k+1)}, \dots, A_{iN}^{(k+1)} \right], i=1,\dots, N$,  
\begin{equation}
	\small
	\begin{aligned}
		&\mathop{\arg\min}\limits_{A_{i}} \mathcal{L}\left(w^{(k)}, A_{i}, \lambda^{(k)}\right) \\
		= & \mathop{\arg\min}\limits_{A_{i}} \sum_{j=1}^N A_{ij} \\ 
		& + \lambda_i^{(k)}\left( \dot{x}_i - \sum_{m=1}^{M_1} w_{im}^{(k)} F_m(x_i) - \sum_{m=1}^{M_2} w_{im}^{(k)} \sum_{j=1}^N A_{ij}G_m(x_i, x_j) \right) \\ 
		& + \frac{\rho}{2} \left\Vert \dot{x}_i - \sum_{m=1}^{M_1} w_{im}^{(k)} F_m(x_i) - \sum_{m=1}^{M_2} w_{im}^{(k)} \sum_{j=1}^N A_{ij}G_m(x_i, x_j) \right\Vert_2^2. \\
		&\text{s.t.}~A_{ij} \geq 0, j = 1, \dots, N.
	\end{aligned}
	\label{update_A}
\end{equation}

Then, to update $w^{(k+1)}$ in Eq.~(\ref{wk}), independently for $w_{i}^{(k+1)} = \left[ w_{i1}^{(k+1)}, \dots, w_{i(M_1+M_2)}^{(k+1)} \right], i=1,\dots, N$,
\begin{equation}
	\fontsize{7.9pt}{10pt}\selectfont
	\begin{aligned}
		&\mathop{\arg\min}\limits_{w_i} \mathcal{L}\left( w_i, A^{(k+1)}, \lambda^{(k)} \right) \\
		=& \mathop{\arg\min}\limits_{w_i} \sum_{m=1}^{M_1 + M_2} \left\Vert w_{im} \right\Vert_1 \\
		& + \lambda_i^{(k)}\left( \dot{x}_i - \sum_{m=1}^{M_1} w_{im} F_m(x_i) - \sum_{m=1}^{M_2} w_{im} \sum_{j=1}^N A_{ij}^{(k+1)} G_m(x_i, x_j) \right) \\ 
		& + \frac{\rho}{2} \left\Vert \dot{x}_i - \sum_{m=1}^{M_1} w_{im} F_m(x_i) - \sum_{m=1}^{M_2} w_{im} \sum_{j=1}^N A_{ij}^{(k+1)} G_m(x_i, x_j) \right\Vert_2^2.
	\end{aligned}
	\label{update_w}
\end{equation}
To solve the Eq.~(\ref{update_w}), we set $u_i = \left[ u_{i1}, \dots, u_{i(M_1 + M_2)} \right]$, and $u_{im} = \max(0, w_{im}), m = 1, \dots, M_1 + M_2$. And then, we set $v_i = \left[ v_{i1}, \dots, v_{i(M_1 + M_2)} \right]$, and $-v_{im} = \min(0, w_{im}), m = 1, \dots, M_1 + M_2$. Thus, it is easy to obtain, $u_{im} \geq 0, v_{im} \geq 0$, and $w_{im} = \max(0, w_{im}) + \min(0, w_{im}) = u_{im} -v_{im}$. Then, Eq.~(\ref{update_w}) can be rewritten as,
\begin{equation}
	\begin{aligned}
		& \mathop{\arg\min}\limits_{u_i, v_i} \sum_{m=1}^{M_1 + M_2} u_{im} + \sum_{m=1}^{M_1 + M_2} v_{im} \\
		& + \lambda_i^{(k)} \bigg[\dot{x}_i - \sum_{m=1}^{M_1} (u_{im} - v_{im}) F_m(x_i)  \\ 
		&- \sum_{m=1}^{M_2} (u_{im} - v_{im}) \sum_{j=1}^N A_{ij}^{(k+1)} G_m(x_i, x_j) \bigg] \\ 
		& + \frac{\rho}{2} \bigg\Vert\dot{x}_i - \sum_{m=1}^{M_1} (u_{im} - v_{im}) F_m(x_i) \\
		& - \sum_{m=1}^{M_2} (u_{im} - v_{im}) \sum_{j=1}^N A_{ij}^{(k+1)} G_m(x_i, x_j)\bigg\Vert_2^2, \\
		& \text{s.t.}~u_{im} \geq 0, v_{im} \geq 0, m = 1, \dots, M_1 + M_2.
	\end{aligned}
	\label{update_z}
\end{equation}
Then the $l_1$-normalization term in Eq.~(\ref{update_w}) is released, and Eq.~(\ref{update_w}) can be equivalently solved by Eq.~(\ref{update_z}). If $u_i$ and $v_i$ are both obtained, $w_i$ can be ontained by $w_i = u_i - v_i$. It is trivial to see, Eq.~(\ref{update_A}) and Eq.~(\ref{update_z}) are both quadratic programming problem, and they have the same form for $A_i$, $u_i$ and $v_i$. Thus, we can use the same algorithm to solve them. 

Then, to update $\lambda^{(k+1)}$ in Eq.~(\ref{lk}), we can obtain that
\begin{equation}
	\small
	\begin{aligned}
		&\frac{\partial \mathcal{L}(w^{(k+1)}, A^{(k+1)}, \lambda)}{\partial \lambda} \\
		= &\dot{x}_i - \sum_{m=1}^{M_1} w_{im}^{(k+1)} F_m(x_i) - \sum_{m=1}^{M_2} w_{im}^{(k+1)} \sum_{j=1}^N A_{ij}^{(k+1)}G_m(x_i, x_j),
	\end{aligned}
\end{equation}
and we can use it to replace the corresponding term in Eq.~(\ref{lk}). Thus, due to the convexity of Eq.~(\ref{new}), the unique solution can be obtained, and it is the same solution to Eq.~(\ref{original}).


\section{Prediction Experiments}

To test the prediction performance of ASIND algorithm, this work conducts multiple experiments to predict network dynamics over 100 steps, with different networks and different functions on Eq.~(\ref{netdyns}) for universal social systems. 

\subsection{Models of Dynamics on Networks}

Four universal network dynamics are used here to initialize Eq.~(\ref{netdyns}), as shown in Table~\ref{11dyns}. Among them: 

\textbf{Kuramoto} model~\citep{kuramoto2005, kuramoto2016} is used to describe the synchronization between $N$ coupled oscillators through phase coupling. In the model, each oscillator has its own natural frequency, and the coupling term promotes interactions between oscillators, which leads to the collective behavior. Thus, $\omega_i$ is the frequency of oscillator $i$, and $c>0$ is the coupling coefficient.

\textbf{Susceptible-Infected-Susceptible (SIS)} model~\citep{SIS1975, SIS2001} is used to describe the synchronization between $N$ individuals, that the state of individuals switch cyclically between susceptible and infected states. Each individual can recover from the infected state, and later become infected again. The SIS model is widely used to study the epidemic dynamics in populations, thus here $\delta_i > 0$ is recovery rate, and  $\gamma_i \geq 0$ is infection rate.

\textbf{Lotka-Volterra (LV)} model~\citep{LV1970} is used to describe the dynamical evolution between $N$ species competing for shared resources within an ecosystem. The growth of each species is influenced by its own population density and the population density of other competing species, which leads to coexistence and competitive exclusion between natural species. Thus, $\alpha_i > 0 $ and $\theta_i > 0$ control the intrinsic growth rate of species $i$ together, and $\gamma_i$ is the competition coefficient.

\textbf{Michaelis-Menten (MM)} model~\citep{MM2017, MM2019, MM2016Gao} is used to describe the interactions between enzymes and substrates in biological networks, and interpret the signal transmission in metabolic pathways. The model demonstrates the mechanism of steady state behaviors and response characteristics of the biochemical systems. Thus, $h$ is Hill coefficient, which represents the binding strength between the enzyme and its substrate.

\begin{table}[htbp]
	\centering
	\caption{Models of Dynamics on Networks}
	\label{11dyns}
	\setlength{\tabcolsep}{17pt}
	\renewcommand{\arraystretch}{1.4}
	\begin{tabular}{c|cc}
		\hline
		Model & $F(x_i)$ & $G(x_i, x_j)$ \\
		\hline
		Kuramoto & $\omega_i$ & $ \frac{c}{N} \sin(x_j - x_i)$ \\
		SIS & $-\delta_i x_i$ & $ \gamma_i(1 - x_i)x_j$ \\
		LV & $x_i(\alpha_i - \theta_i x_i)$ & $-\gamma_i x_i x_j$ \\
		MM & $-x_i$ & $x_j^h (1 + x_j^h)^{-1}$ \\
		\hline
	\end{tabular}
\end{table}

Moreover, the adjacency $A$ in Eq.~(\ref{netdyns}) is initialized with Erd\H{o}s-R\'{e}nyi (ER) graph~\citep{ergraph1959}, Watts-Strogatz (WS) network~\citep{smallworld1998nature} and Barab\'{a}si-Albert (BA) network~\citep{scalefree1999science}, respectively. In ER graph, the connection probability $p=0.1$. In WS network, the average degree $\left\langle k \right\rangle=4$ and the rewiring probability $p=0.1$. In BA network, the three control parameters $\alpha=0.41, \beta=0.54, \gamma=0.05$. The network size is 16.

\begin{table*}[t]
	\centering
	\caption{100-steps Prediction performance evaluation }
	\setlength{\tabcolsep}{9pt}
	\renewcommand{\arraystretch}{1.1}
	\label{performacne}
	\begin{tabular}{c|c|cc|cc|cc}
		\hline
		\multicolumn{2}{c|}{~} & \multicolumn{2}{c|}{Erd\H{o}s-R\'{e}nyi} & \multicolumn{2}{c|}{Watts-Strogatz} & \multicolumn{2}{c}{Barab\'{a}si-Albert} \\
		\cline{3-8}
		\multicolumn{2}{c|}{~} & RMSE & MAPE & RMSE & MAPE & RMSE & MAPE \\
		\hline
		\multirow{2}{*}{Kuramoto} & SINDy & 0.1161 & 13.42\% & 0.0041 & 0.05\% & 0.1183 & 1.96\% \\
		& ASIND & 0.2355 & \textbf{3.37\%} & 0.0966 & 1.26\% & \textbf{0.0327} & \textbf{1.01\%} \\ 
		\hline
		\multirow{2}{*}{SIS} & SINDy & 6686.04 & 315900.43\% & 272147.23 & 11167042.65\% & 9543.43 & 1005056.69\% \\
		& ASIND & \textbf{0.0005} &  \textbf{0.36\%} & \textbf{0.0011} & \textbf{0.33\%} & \textbf{0.0002} & \textbf{0.24\%} \\ 
		\hline
		\multirow{2}{*}{LV} & SINDy & 161229.61 & 2428084.23\% & 59600665.46 & 2284056928.65\% & 91471578.44 & 7010535232.61\% \\
		& ASIND & \textbf{0.0007} & \textbf{0.05\%} & \textbf{0.0021} &  \textbf{0.09\%} & \textbf{0.0017} & \textbf{0.25\%} \\ 
		\hline
		\multirow{2}{*}{MM} & SINDy & 4952779.64 & 152395650.85\% & 435611.43 & 22165586.15\% & 3076443.27 & 173795298.93\% \\
		& ASIND & \textbf{0.0006} & \textbf{0.17\%} & \textbf{0.0012} & \textbf{0.33\%} & \textbf{0.0007} & \textbf{0.23\%} \\ 
		\hline
	\end{tabular}
\end{table*}

\subsection{Baseline}


The prediction performance of ASIND is compared to that of the baseline \textbf{SINDy} algorithm~\citep{SINDy2016PNAS}. Note that, the two algorithms are both conducted without any knowledge, and can be used for universal network dynamics identification.

SINDy is proposed to sparsely identify the governing equations in nonlinear dynamical systems. The algorithm uses a group of basis functions to make a basis matrix from observation data, and then identifies the weights of the basis functions by $l_0$-normalization optimization, to reconstruct the nonlinear dynamics finally. Here, the basis matrix is mainly composed of polynomial basis up to second order.

\subsection{Evaluation Indices}

To compare the prediction performance of ASIND algorithm with that of the other baseline, root mean square error (RMSE) and mean absolute percentage error (MAPE) are introduced to evaluate prediction performance, as follows:
\begin{equation}
	\begin{aligned}
		\text{RMSE} &= \sqrt{\frac{1}{T} \sum_{t=1}^T \left(\hat{x}_t - x_t \right)^2}, \\
		\text{MAPE} &= \frac{1}{T} \sum_{t=1}^T \left| \frac{\hat{x}_t - x_t}{x_t}\right| \times 100\%, 
	\end{aligned}
\end{equation}
where $x_t\in \mathbb{R}^N$ is the observation activities of the $N$ nodes at time $t$, and $\hat{x}_t$ is the corresponding prediction.   

Moreover, to quantify and measure the identifiability of network $A$ in Eq.~(\ref{netdyns}), Jaccard index is introduced as follows:
\begin{equation}
	J(A, \hat{A}) = \frac{A \cap \hat{A}}{A \cup \hat{A}} \times 100\%,
\end{equation}
where $\hat{A}$ is the estimation of $A$. Note that, all elements of $A$ and $\hat{A}$ are converted to zero (i.e., 0) and non-zero elements (i.e., 1) when calculate Jaccard index.

\subsection{Prediction Performance}

As shown in Table~\ref{performacne}, ASIND algorithm achieves the state-of-the-art performance on the task of 100-steps prediction, and the error indices are significantly low, especially for the network dynamics including SIS model, LV model and MM model. This means the prediction results of ASIND are close to the ground truths during long-term prediction, and the network dynamics are identified accurately.

Meanwhile, the prediction results of SINDy significantly deviate from the ground truths for 100-steps prediction, that means SINDy cannot identify the network dynamics accurately. However, it is found that SINDy perform well on the Kuramoto model. A possible reason is that high-order polynomial bases can well approximate the sine function, but hardly approximate the complex functions in other network dynamical models.

\subsection{Weak Identifiability of Network Structure}

As shown in Table~\ref{jaccard}, the network structure $A$ in Eq.~(\ref{netdyns}) is weakly identifiable. The ASIND algorithm can be trained on the observation data and then predict 100-steps evaluation accurately, but the Jaccard index $J(A, \hat{A})$ is still low for different network dynamics. This means, the reconstructed networks and the ground-truth networks are equivalent to represent the network dynamics, and that makes the ground-truth networks difficult to be identified from the observation data.

\begin{table}[htbp]
	\centering
	\caption{The Jaccard index between the reconstructed network and the ground truth}
	\setlength{\tabcolsep}{6pt}
	\renewcommand{\arraystretch}{1.4}
	\label{jaccard}
	\begin{tabular}{c|c|c|c}
		\hline
		$J(A, \hat{A})$ & Erd\H{o}s-R\'{e}nyi & Watts-Strogatz & Barab\'{a}si-Albert \\
		\hline
		Kuramoto & 17.86\% &  15.38\%   & 15.12\% \\
		\hline
		SIS &  18.75\%  &  7.69\% & 2.7\% \\ 
		\hline
		LV & 5.88\%  &  9.52\%  & 3.08\% \\ 
		\hline
		MM & 2.44\% &  4.88\%  & 5.66\% \\
		\hline
	\end{tabular}
\end{table}

\section{Conclusion}

ASIND algorithm is proposed here to sparsely identify the network dynamics without the knowledge of $F$, $G$ and $A$. Experiments are conducted to show the state-of-the-art identification and 100-steps prediction performance, compared to the popular SINDy algorithm. Moreover, the results also show that the network structure $A$ has weak identifiability, that means different networks can generate highly similar trajectories of network dynamics. 

Of course, the experiments here are oversimplification for a fair comparison with SINDy, e.g., network size $N=16$, which is not in line with the real-world systems. In fact, the maximum network size ASIND can support is $N=100$. If the number goes beyond this point, the performance gradually declines.

\bibliography{ifacconf}             
                                                   







\end{document}